\documentstyle[12pt]{article}

\input epsf
\begin{document}
\baselineskip=20.5pt

\def\beqra{\begin{eqnarray}} \def\eeqra{\end{eqnarray}}
\def\beqast{\begin{eqnarray*}}
\def\eeqast{\end{eqnarray*}}
\def\beq{\begin{equation}}      \def\eeq{\end{equation}}
\def\be{\begin{enumerate}}   \def\ee{\end{enumerate}}

%title page
\def\fnote#1#2{\begingroup\def\thefootnote{#1}\footnote{
#2}
\addtocounter
{footnote}{-1}\endgroup}

\def\itp#1#2{\hfill{NSF-ITP-{#1}-{#2}}}

\def\gam{\gamma}
\def\Gam{\Gamma}
\def\la{\lambda}
\def\eps{\epsilon}
\def\La{\Lambda}
\def\si{\sigma}
\def\Si{\Sigma}
\def\al{\alpha}
\def\Tha{\Theta}
\def\tha{\theta}
\def\vphi{\varphi}
\def\del{\delta}
\def\Del{\Delta}
\def\ab{\alpha\beta}
\def\om{\omega}
\def\Om{\Omega}
\def\mn{\mu\nu}
\def\mun{^{\mu}{}_{\nu}}
\def\kap{\kappa}
\def\rsi{\rho\sigma}
\def\beal{\beta\alpha}

\def\til{\tilde}
\def\htil{\tilde{H}}
\def\rta{\rightarrow}
\def\eqv{\equiv}
\def\nab{\nabla}
\def\pa{\partial}
\def\sit{\tilde\sigma}
\def\ul{\underline}
\def\indt{\parindent2.5em}
\def\nd{\noindent}

\def\rsi{\rho\sigma}
\def\beal{\beta\alpha}

        % calligraphic
\def\caa{{\cal A}}
\def\cb{{\cal B}}
\def\cac{{\cal C}}
\def\cd{{\cal D}}
\def\ce{{\cal E}}
\def\cf{{\cal F}}
\def\cg{{\cal G}}
\def\cah{{\cal H}}
\def\ci{{\cal I}}
\def\cj{{\cal{J}}}
\def\ck{{\cal K}}
\def\cl{{\cal L}}
\def\cm{{\cal M}}
\def\cn{{\cal N}}
\def\cO{{\cal O}}
\def\cp{{\cal P}}
\def\car{{\cal R}}
\def\cs{{\cal S}}
\def\ct{{\cal{T}}}
\def\cu{{\ca{U}}}
\def\cv{{\cal{V}}}
\def\cw{{\cal{W}}}
\def\cx{{\cal{X}}}
\def\cy{{\cal{Y}}}
\def\cz{{\cal{Z}}}

        % nots
\def\raisenot{\raise .5mm\hbox{/}}
\def\nota{\ \hbox{{$a$}\kern-.49em\hbox{/}}}
\def\notA{\hbox{{$A$}\kern-.54em\hbox{\raisenot}}}
\def\notb{\ \hbox{{$b$}\kern-.47em\hbox{/}}}
\def\notB{\ \hbox{{$B$}\kern-.60em\hbox{\raisenot}}}
\def\notc{\ \hbox{{$c$}\kern-.45em\hbox{/}}}
\def\notd{\ \hbox{{$d$}\kern-.53em\hbox{/}}}
\def\notbd{\ \hbox{{$D$}\kern-.61em\hbox{\raisenot}}} %big D
\def\note{\ \hbox{{$e$}\kern-.47em\hbox{/}}}
\def\notk{\ \hbox{{$k$}\kern-.51em\hbox{/}}}
\def\notp{\ \hbox{{$p$}\kern-.43em\hbox{/}}}
\def\notq{\ \hbox{{$q$}\kern-.47em\hbox{/}}}
\def\notW{\ \hbox{{$W$}\kern-.75em\hbox{\raisenot}}}
\def\notz{\ \hbox{{$Z$}\kern-.61em\hbox{\raisenot}}}
\def\notpa{\hbox{{$\partial$}\kern-.54em\hbox{\raisenot}}}

\def\fo{\hbox{{1}\kern-.25em\hbox{l}}}  %raised one
\def\rf#1{$^{#1}$}
\def\bx{\Box}
\def\tr{{\rm Tr}}
\def\rmtr{{\rm tr}}
\def\dgg{\dagger}

\def\lag{\langle}
\def\rag{\rangle}
\def\bmid{\big|}
\def\pw{P\left(w\right)}

\def\vlap{\overrightarrow{\La p}} %overrightarrow
\def\lrta{\longrightarrow}
\def\lrar{\raisebox{.8ex}{$\longrightarrow$}}
\def\rlarw{\longleftarrow\!\!\!\!\!\!\!\!\!\!\!\lrar}

\def\llra{\relbar\joinrel\longrightarrow}     %THIS IS LONG
\def\mapright#1{\smash{\mathop{\llra}\limits_{#1}}}
\def\mapup#1{\smash{\mathop{\llra}\limits^{#1}}}
\def\asymptotic{{_{\stackrel{\displaystyle\longrightarrow}
{x\rightarrow\pm\infty}}\,\, }} %x goes to plus minus infinity, display sty.
\def\asymptext{\raisebox{.6ex}{${_{\stackrel{\displaystyle\longrightarrow}
{x\rightarrow\pm\infty}}\,\, }$}} %x goes to plus minus infinity,

\def\7#1#2{\mathop{\null#2}\limits^{#1}}   % puts #1 atop #2
\def\5#1#2{\mathop{\null#2}\limits_{#1}}   % puts #1 beneath #2
\def\too#1{\stackrel{#1}{\to}}
\def\tooo#1{\stackrel{#1}{\longleftarrow}}
\def\nout{{\rm in \atop out}}

\def\one{\raisebox{.5ex}{1}}
\def\BM#1{\mbox{\boldmath{$#1$}}}

\def\ltsim{\matrix{<\cr\noalign{\vskip-7pt}\sim\cr}}
\def\gtsim{\matrix{>\cr\noalign{\vskip-7pt}\sim\cr}}
\def\haf{\frac{1}{2}}

%       pictures

\def\place#1#2#3{\vbox to0pt{\kern-\parskip\kern-7pt
                             \kern-#2truein\hbox{\kern#1truein #3}
                             \vss}\nointerlineskip}

\def\illustration #1 by #2 (#3){\vbox to #2{\hrule width #1
height 0pt
depth
0pt
                                       \vfill\special{illustration #3}}}

\def\scaledillustration #1 by #2 (#3 scaled #4){{\dimen0=#1
\dimen1=#2
           \divide\dimen0 by 1000 \multiply\dimen0 by #4
            \divide\dimen1 by 1000 \multiply\dimen1 by #4
            \illustration \dimen0 by \dimen1 (#3 scaled #4)}}

\def\ON{{\cal O}(N)}
\def\UN{{\cal U}(N)}
\def\bdPh{\mbox{\boldmath{$\dot{\!\Phi}$}}}
\def\bPh{\mbox{\boldmath{$\Phi$}}}
\def\bPhs{\bPh^2}
\def\sef{S_{eff}[\sigma,\pi]}
\def\sigx{\sigma(x)}
\def\pix{\pi(x)}
\def\bph{\mbox{\boldmath{$\phi$}}}
\def\bphs{\bph^2}
\def\ex{\BM{x}}
\def\exs{\ex^2}
\def\xdot{\dot{\!\ex}}
\def\y{\BM{y}}
\def\ys{\y^2}
\def\ydot{\dot{\!\y}}
\def\pat{\pa_t}
\def\pax{\pa_x}

\renewcommand{\theequation}{\arabic{equation}}

%\today

\itp{97}{115}\\

\vspace*{.3in}
\begin{center}
 \large{\bf Spectral Curves of Non-Hermitean Hamiltonians}

\vspace{36pt}
{J. Feinberg$^{a}$ \& A. Zee$^{b}$}
\end{center}

\vskip 2mm
\begin{center}$^{a)}$ Department of Physics, \\
Oranim-University of Haifa, Tivon 36006, Israel\\

 $^{b)}$~{Institute for Theoretical Physics,}\\ 
{University of California, Santa Barbara, CA 93106, USA}
\vspace{.6cm}

\end{center}

\begin{minipage}{5.3in}
{\abstract~~~~~ Recent analytical and numerical work have shown that the spectrum of the random non-hermitean Hamiltonian on a ring which models the physics of vortex line pinning in superconductors is one dimensional. In the maximally non-hermitean limit, we give a simple ``one-line" proof of this feature. We then study the spectral curves for various distributions of the random site energies. We find that a critical transition occurs when the average of the logarithm of the random site energy squared vanishes. For a large class of probability distributions of the site energies, we find that as the randomness increases the energy $E_\ast$ at which the localization-delocalization transition occurs increases, reaches a maximum, and then decreases. The Cauchy distribution studied previously in the literature does not have this generic behavior. We determine $\gam_{c1}$, the critical value of the randomness at which ``wings" first appear in the energy spectrum. For distributions, such as Cauchy, with infinitely long tails, we show that $\gam_{c1} ~=~ 0^+$. We determine the density of eigenvalues on the wings for any probability distribution. We show that the localization length on the wings diverge generically as $L(E) ~\sim~{1 \over \mid E-E_\ast \mid}$ as $E$ approaches $E_\ast$. These results are all obtained in the maximally non-hermitean limit but for a generic class of probability distributions of the random site energies.}
\end{minipage}

\vspace{48pt}
\vfill
\pagebreak

\setcounter{page}{1}

\section{Introduction}

Recently, there has been a great deal of work on non-hermitean matrices \cite{recent}, operators \cite{mw}, \cite{ns}, and Hamiltonians \cite{hatano}. To make this paper as self-contained as possible, we will begin with a brief review of the material relevant for our discussion. Non-hermitean operators appear in a  broad class of problems in statistical physics describing diffusion and growth \cite{private} exemplified by the equation

\beq\label{dif}
{dP_j \over dt} ~=~ w_{j+1,j} ~ P_{j+1} ~+~ w_{j-1,j} ~ P_{j-1} ~+~ w_{j,j} P_j
\eeq
with $P_j$ representing for example the concentration on site $j$ or the population of species $j$. The $w$'s denote various transition probabilities and growth rates. Writing (\ref{dif}) as

\beq\label{dif2}
{d \stackrel{\rightarrow}{P} \over dt} ~=~ L \stackrel{\rightarrow}{P}
\eeq
we see that the operator $L$ is in general non-hermitean. 

A non-hermitean Hamiltonian inspired by the problem of vortex line pinning in superconductors has attracted the attention of a number of authors \cite{efetov, feinzee, zahed, beenakker, zee, bz}. In one version, we are to study the eigenvalue problem

\beq\label{eig}
\sum_j~H_{ij} \psi_j ~=~ {t \over 2} \left(e^h ~ \psi_{i+1} ~+~ e^{-h}~\psi_{i-1}\right) ~+~ w_i \psi_i ~=~ E\psi_i, ~i,j ~=~ 1, \ldots, N
\eeq
with the periodic identification $i+N\equiv i$ of site indices. Here the real numbers $w_i$ are drawn independently from some probability distribution $P\left(w\right)$ (which we will henceforth take to be even for simplicity.)
This Hamiltonian describes a particle hopping on a ring, with its clockwise hopping amplitude different from its counter-clockwise hopping amplitude. On each site there is a random potential. The number of sites $N$ is understood to be tending to infinity. We will also take $t$ and $h$ to be positive for definiteness. (The hopping amplitude $t$ can be scaled to $2$ for instance but we will keep it for later convenience.) Note that the Hamiltonian is non-hermitean for $h$ non-zero and thus has complex eigenvalues. It 
is represented by a real non-symmetric matrix, with the reality implying 
that if $E$ is an eigenvalue, then $E^*$ is also an eigenvalue. For $\pw$ even, for each particular realization $\{w_i\}$ of the random site energies, the realization $\{-w_i\}$ is equally likely to occur, and thus the spectrum of $H$ averaged with $\pw$ has the additional symmetry $E ~\rta~ -E$. This eigenvalue problem is clearly a special case of (\ref{dif}).

Without non-hermiticity ($h=0$), all eigenvalues are of course real, and Anderson and collaborators \cite{anderson} showed that all states are localized. Without impurities ($w_i=0$), Bloch told us that
the Hamiltonian is immediately solvable with the eigenvalues
\beq\label{spectrum}
E_n = t~{\rm cos}~( {2\pi n\over N} - ih)\,, \quad (n = 0, 1, \cdots, N-1)\,, 
\eeq
tracing out an ellipse. The corresponding wave functions $\psi^{(n)}_j\sim {\rm exp}~2\pi i n j/N$ are obviously extended. We are to study what happens in the presence of both non-hermiticity and the impurities.
\begin{figure}[htbp]
\epsfxsize=4in
\begin{center}
\leavevmode
\epsfbox{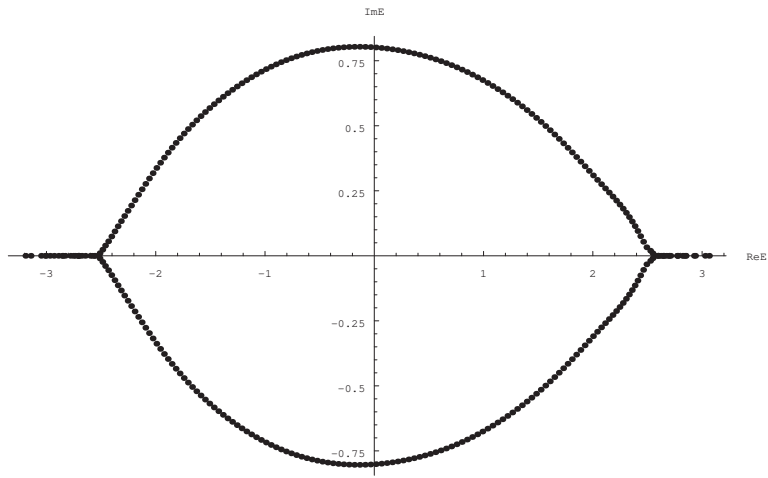}
\end{center}
\caption{The eigenvalues of one particular realization of the Hamiltonian in (3) with $N=300$, $t=2$, $h=0.5$ and with the site energies drawn from the box distribution (13) with $V=1.8$.}
\label{fig1}
\end{figure}

With impurities present, two ``wings," a right wing and a left wing, emerge out of the two ends of the ellipse. (See figure 1.) Some eigenvalues are now real. Evidently, the ``forks" where the two wings emerge out of the ``ellipse-like" curve  represent a non-perturbative effect, and cannot be obtained by treating the impurities perturbatively. For $\pw$ even, we can focus on the right wing in light of the symmetry mentioned above.

Le Doussal \cite{ld} and Hatano and Nelson \cite{hatano} emphasized that $H$ has a special property, namely that 
by a (non-unitary) gauge transformation \cite{dg, mf} $\psi_j ~=~ e^{-jh}~\tilde{\psi}_j$, all the non-hermiticity can be concentrated on one arbitrarily chosen bond, on which the hopping amplitudes are ${t \over 2}~e^{\pm~Nh}$, without changing the spectrum. Indeed, performing this transformation on (\ref{eig}) we obtain the eigenvalue problem $\tilde H~ \tilde\psi ~=~ E \tilde\psi$ with

\beqra\label{H1}
\tilde{H}_{ij} &= &{t \over 2} \left(\delta_{i+1,j} ~+~ \delta_{i-1,j}\right) ~+~ w_j ~ \delta_{ij} \nonumber \\ 
\tilde{H}_{N1} &= &{t \over 2} e^{Nh} , \nonumber \\
\tilde{H}_{1N} &= &{t \over 2}e^{-Nh}
\eeqra
The non-hermiticity has been moved to be on only one particular link, which in (\ref{H1}) is between sites $1$ and $N$, but it is clear that by a suitable transformation we can put the hopping amplitude ${t \over 2}~e^{\pm~ Nh}$ on any link. 

This leads to an extraordinarily simple argument \cite{hatano} that the states
corresponding to complex eigenvalues are extended, that is, delocalized.
Assume that an eigenstate of $\tilde{\psi}$ of $\tilde{H}$ is localized around some site $j$. We can always gauge the non-hermiticity to a link which is located as
far away from the site $j$ as possible, where $|\tilde{\psi}|$ is exponentially small.   Thus, if 
we cut the ring open at that link, the effect on the Schr\"odinger equation would be exponentially small, if the localization length of the state is smaller than $h^{-1}$. The state does not know that the Hamiltonian is non-hermitean and so its eigenvalue must be real.  
Transforming back to $H$, we have a localized state whose eigenvalue is real. On the other hand, if $h$ is larger than the inverse localization length of this state, we no longer have a normalized eigenstate of $H$. Hatano and Nelson \cite{hatano} argued that such states of $\tilde{H}$ and extended states of $\tilde{H}$ correspond to states of $H$ with complex eigenvalues. They thus concluded that the states of $H$ corresponding to complex eigenvalues are extended, that is, delocalized.

Remarkably, non-hermitean localization theory is simpler in this respect
than the standard hermitean localization theory of Anderson and others \cite{anderson}. To understand the localization transition, we have to study only the energy spectrum of $H$.

One way to think about this problem is to imagine starting with $H_0$ (that is, $H$ with $\{w_i\}$ set to zero) and then increasing the randomness (measured by a parameter $\gam$, say) and ask if there is a critical strength of randomness $\gam_{c1}$ at which localized states first appear. There may also be another critical strength $\gam_c$ at which all the extended states become localized. Obviously another way of thinking about the problem is to start with the hermitean random problem $(h=0)$ in which all states are localized. We then increase the non-hermiticity and ask about the critical strength $h_c$ of the non-hermiticity at which extended states first appear. Similarly, we can ask, for given $h$ and $\gam$, the critical hopping amplitude $t_c$ needed for the states to become delocalized. 

As we have mentioned, a number of authors have considered (\ref{eig}) . In this paper, we will refer extensively to \cite{zee} and \cite{bz} which we will denote by FZ and BZ respectively. Here, as in FZ and BZ, the emphasis is to understand analytically the essential physics involved.

We will now describe some of the results from FZ and BZ, partly to fix the framework for our subsequent discussion. As is standard, we are to study the Green's function

\beq\label{green}
G(z) = {1 \over z-H}
\eeq
The averaged density of eigenvalues, defined by 

\beq\label{master}
\rho(x,y) ~\equiv~ \langle {1\over N} \sum_i \delta(x-{\rm Re} E_i)~ \delta (y- {\rm Im}
E_i)\rangle \,,
\eeq
is then obtained (upon recalling the identity ${\partial \over \partial z^\ast} ~ {1 \over z} ~=~ \pi \delta(x) ~\delta(y)$) as

\beq\label{master2}
\rho\left(x,y\right) ~=~ {1 \over \pi} {\partial \over \partial z^\ast} \overline{G\left(z\right)}
\eeq
with $z=x+iy$. Here we define

\beq
\overline{G\left(z\right)} \equiv \lag \frac{1}{N}~ {\rm tr}~ G \left(z\right) \rag
\eeq
and $< \ldots >$ denotes, as usual, averaging with respect to the probability ensemble from which $H$ is drawn. A small subtlety is that while $G(z)$ is ostensibly a function of $z$ only, $ \lag G(z) \rag$ depends on both $z$ and $z^{\ast}$. A careful discussion is given in \cite{feinzee}. It is also useful to define the Green's function $G_0(z) = {1 \over z-H_0}$ in the absence of impurities and the corresponding $\overline{G_0(z)} \equiv {1 \over N} ~{\rm tr} ~ {1 \over z - H_0}$. For the spectrum in (\ref{spectrum}), we can determine $\overline{G_0(z)}$ explicitly \cite{bz}.

In the literature \cite{recent}, the potential

\beq\label{pot}
\Phi(z,z^\ast) ~\equiv~ {1 \over 2N}~ \lag ~{\rm Tr} ~log~(z-H)~(z^\ast-H^\dgg) \rag
\eeq 
has also been introduced. We see that the density of eigenvalues is given by

\beq\label{dens}
\rho(x,y) ~=~ {2 \over \pi} ~ {\partial^2 \over \partial z~ \partial z^\ast}~ \Phi(z,z^\ast)
\eeq
which can be interpreted as the charge density generating the two-dimensional electrostatics potential $\Phi(z,z^\ast)$. See \cite{feinzee} for a careful discussion.

The following probability distributions were studied in FZ and BZ: the sign distribution

\beq\label{pw1}
P(w) = {1\over 2} [\delta (w-r) + \delta (w+r)]
\eeq
with some scale $r$, the box distribution 
\beq\label{box}
P(w) = {1\over 2V}~\theta (V^2-w^2)
\eeq
obtained by ``smearing" the sign distribution, and the Cauchy distribution
\beq\label{Cauchy}
P(w) = {\gam\over \pi} {1\over w^2+\gam^2}\,,
\eeq
with its long tails extending to infinity. 

We might also wish to consider the effect of diluting the impurities by setting randomly some fraction $d$ of the $w_k$'s to zero. In other words, given $P(w)$ we can consider

\beq\label{dil}
P_d(w) ~=~ d~ \delta (w) ~+~ (1-d) ~ P(w)
\eeq
with $0 \leq d \leq 1$. As $d$ increases, there should be more extended states. 

In BZ, a general formalism for studying this problem was established. A formula for the Green's function

\beq\label{together}
G~=~ \left[1-\sum_k\left(\frac{\upsilon_k G_0 P_k}{1+\upsilon_k G_0 P_k} \right) \right]^{-1} ~ G_0 ~=~\left[1-\sum_k \frac{\upsilon_k G_0 P_k}{1+\upsilon_k \overline{G_0}} \right]^{-1} ~ G_0
\eeq
was obtained.
Here we have defined the projection operator
\beq
P_k ~=~ \mid k \rag \lag k \mid
\eeq
onto the site $k$ and the repeated scattering amplitude on the impurity potential at site $k$

\beq\label{summation}
\upsilon_k ~\equiv~ {w_k \over 1-w_k \lag k \mid G_0 \mid k \rag}
\eeq

The reader is invited to average the trace of (\ref{together}) with his or her favorite $\pw$, and thence to obtain the density of eigenvalues $\rho(x,y)$ by (\ref{master2}). For an arbitrary $P(w)$ it is presumably not possible to average (\ref{together}) explicitly and obtain $\lag G(z) \rag$ in closed form, not any more than it is possible to obtain $\lag G(z)\rag$ in closed form for the hermitean problem. Indeed, (\ref{together}) was derived by only assuming translation invariance for $H_0$, whether hermitean or non-hermitean. It is, however, always possible to expand (\ref{together}) to any desired powers of $\upsilon$ and average. Some results are given in BZ.

Remarkably, explicit results can be given for the Cauchy distribution. These results were also obtained independently by Goldsheid and Khoruzhenko \cite{gk}. Here we state the results from BZ.

The density of eigenvalues consists of two arcs and two wings. Away from the real axis we have

\beq\label{M}
\rho(x,y) ~=~ \rho_0(x,y + \gamma) ~ \theta(y) ~+~ \rho_0(x, y- \gamma) ~ \theta(-y)
\eeq
with $\rho_0$ the eigenvalue density corresponding to $\overline{G_0(z)}$.
In other words, the two arcs of the ellipse are pushed towards the real axis by a distance $\gamma$. On the two wings, the eigenvalue density is

\beq\label{N}
\rho_{wing}(x,y) ~=~ {1 \over \sqrt{2} \pi} ~ \delta (y) ~ \theta (x^2-x^2_{min} (\gamma)) ~ {\sqrt{\gamma^2-x^2+t^2 ~+~ B(x,\gamma,t)} \over B(x, \gamma, t)}
\eeq
where

\beq
B(x, \gamma, t) ~\equiv~ \sqrt{(x^2 + \gamma^2)^2 ~+~ 2t^2(\gamma^2-x^2) ~+~ t^4}
\eeq
and

\beq\label{conv}
x_{min}(\gamma) ~\equiv~ (\sqrt{(t ~{\rm sinh}~ h)^2 ~-~ \gamma^2})~ {\rm tanh}~ h
\eeq
The critical value of $\gamma$ at which the arcs disappear is given by 

\beq\label{critical}
\gamma_c ~=~ t ~{\rm sinh}~ h
\eeq
All states are now localized.

In this model, $\gam_{c1}~$ is $~0^+$. The wings appear as soon as $\gam ~>~ 0$. We see that for insufficient non-hermiticity, $h~<~h_c$ where

\beq
h_c ~=~ {\rm sinh}^{-1} ~ ({\gam \over t}) \,,
\eeq
all states are localized. Note that in accordance with Anderson et al \cite{anderson}, for any finite non-zero $\gam$, no matter how small, there are localized states for $h$ small enough. Trivially, we can express (\ref{critical}) in yet another way, by saying that for a given non-hermiticity $h$ and site randomness $\gam$, we need the hopping to exceed a critical strength

\beq\label{critstrength}
t_c ~=~ {\gam \over {\rm sinh} ~h}
\eeq
before states become delocalized.

For $\pw$ other than Cauchy, we are not able to obtain $\rho(x,y)$ explicitly.

In FZ, it was pointed out the the problem (\ref{eig}) simplifies in the maximally non-hermitean or ``one way" limit, in which the parameters in (\ref{eig}) are allowed to tend to the (maximally non-hermitean)  
limit $h\rightarrow\infty$ and $t\rightarrow 0$ such that 
\beq\label{thlim}
{t \over 2}~e^h\rightarrow 1
\eeq
The particle in (\ref{eig}) can only hop one way. The spectrum (\ref{spectrum}) changes into 
\beq\label{circle}
E_n = {\rm exp}~{2\pi i n\over N}\,, \quad (n = 0, 1, \cdots, N-1)\,,
\eeq
and the ellipse associated with (\ref{spectrum}) expands into the unit 
circle. In BZ, bounds on the domain over which the density of eigenvalues is non-zero were obtained for arbitrary $P(w)$'s with bounded support. Furthermore, it was shown that for the sign distribution the original unit circle spectrum of the 
deterministic $H_0$ is distorted by randomness into the curve 

\beq\label{curve}
z^2=r^2 + e^{i\theta}\,,\quad 0\leq\theta<2\pi
\eeq
in the complex $z$ plane. (See figure 2.) Clearly, $r_c=1$ is a critical value of $r$. For $r<1$, the curve (\ref{curve}) is connected, enclosing a dumb-bell region free of energy eigenvalues. At $r=r_c$, the curve traces out a figure eight. For $r>1$ it breaks into two disjoint symmetric lobes that are located to the right and to the left of the imaginary axis.
\begin{figure}[htbp]
\epsfxsize=4in
\begin{center}
\leavevmode
\epsfbox{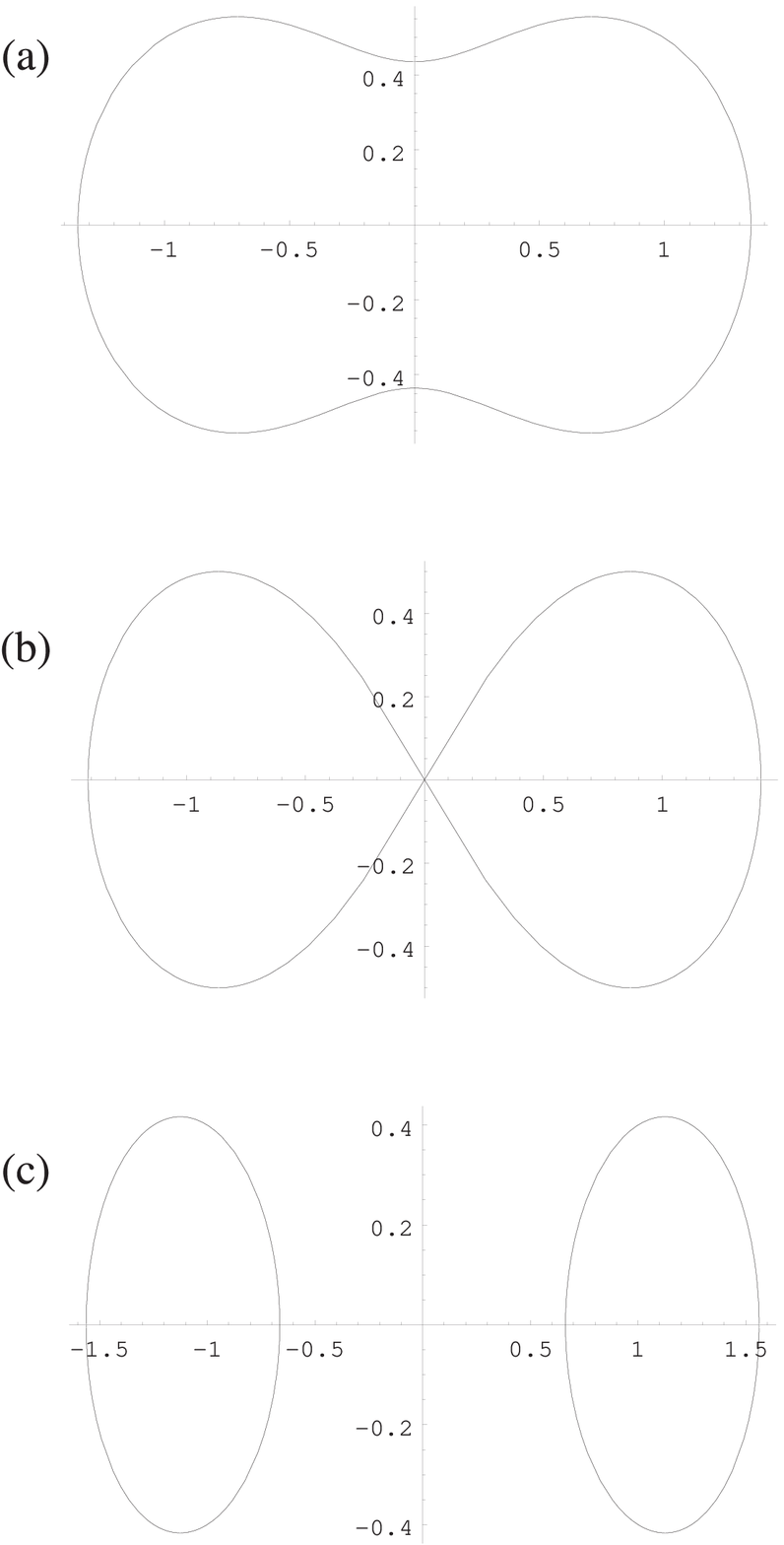}
\end{center}
\caption{The spectral curve for the sign model, for (a) $r=0.9$, (b) $r=1$, and (c) $r=1.2$.}
\label{fig2a}
\end{figure}

Incidentally, it is instructive to understand this phase transition in the language of (\ref{dif}) and in particular the appearance of a zero eigenvalue at $r=1$. In the one way limit we have for $r=1$

\beq\label{dif3}
{dP_j \over dt} ~=~ P_{j+1} ~+~ w_jP_j
\eeq
A static solution ${dP_j \over dt} = 0$ is obtained with $P_{j+1} ~=~ -w_jP_j$. Since $w_j ~=~ \pm 1$ occur with equal probability, we have a 50\% chance of matching the boundary condition $P_{N+1} = P_1$. Thus, we have a zero eigenvalue in the averaged density of eigenvalues and hence a static state.

Here we should mention that, strictly speaking, it is not clear that the simple argument mentioned above showing that localized states have real eigenvalues and that complex eigenvalues imply extended states would still hold in the one way limit. The spectrum should certainly behave smoothly since after all we are simply letting the clockwise hopping amplitude go to $1$ and the anti-clockwise hopping amplitude go to $0$. We have checked this numerically and can easily check this in the explicitly solvable Cauchy model. We have studied in FZ and BZ the localization properties of various models in the simplifying limit of having only one impurity present, and showed analytically that the states on the wings are localized. Indeed, we found that the localization length $L(E)$ diverges as

\beq\label{locz}
L(E) ~\sim~ {1 \over log ~\mid {E \over E_\ast} \mid} ~\sim~ {1 \over \mid E-E_\ast \mid}
\eeq
as $E$ varies along the wings and approaches the points $\pm E_\ast$ at which the wings are attached to the arcs of complex eigenvalues. We conjectured in BZ that the behavior in (\ref{locz}) is generic. We have verified numerically that in the one way limit the states on the wings are indeed localized. We will thus take as our working assumption that the argument mentioned above continues to hold in the one way limit, and hence refer to $E_\ast$ as the energy of the localization-delocalization transition in the one way limit as well as in the more general two way case. In the next section, we give a simple proof that this assumption is indeed correct.

\pagebreak
\section{Spectral Curves of Non-Hermitean \mbox{Hamiltonians}}

In the explicit solutions mentioned here, the eigenvalue density $\rho(x,y)$ traces out a one-dimensional curve. This feature also appears to hold in various numerical studies of $\rho(x,y)$. Recently, Goldsheid and Khoruzhenko \cite{gk} (to be referred to as GK) gave a rather nice proof of this fact. We will describe a simplified version of their proof in the appendix. (We mention in passing that the spectra of some of the random hopping models studied in FZ are explicitly (and interestingly) two dimensional, but it can be readily checked that these models violate the assumptions made in GK.) 

We now give, in the maximally non-hermitean limit, a simple ``one-line" proof that the spectrum of $H$ is one dimensional. In FZ it was found by direct evaluation that in the maximally non-hermitean limit the determinant of $(z-H)$ simplifies drastically to (we take $N$ to be even for definiteness)

\beq\label{det}
{\rm det}~(z-H) = \left(\prod_{i=1}^N (z-w_i)\right) -1\,
\eeq
(For arbitrary $t$ and $h$ the corresponding formula is considerably more complicated.) Note that (\ref{det}) is completely symmetric in the $\{w_i\}$, and thus, for a given set of site energies it is independent of the way the impurities are 
arranged along the ring. (Incidentally, this property is destroyed if we include a next nearest neighbor hopping $H_{i,i+2} ~=~ s$ in addition to $H_{i, i+1} ~=~ 1$, and thus is not solely a consequence of the particle hopping one way.) 

For an alternative derivation of (\ref{det}) let us write down Schr\"odinger's equation in the maximally non-hermitean limit

\beq\label{sch}
(E-w_j) ~ \psi_j ~=~ \psi_{j+1}
\eeq
Starting with $\psi_1~=~\psi_{N+1}$, we have $\psi_N ~=~ (E-w_N)^{-1}~\psi_1$, $\psi_{N-1} ~=~ \prod\limits_{j=N-1}^N ~(E-w_j)^{-1}~\psi_1$, and so on, until finally $\psi_1 ~=~ \prod\limits_{j=1}^N ~(E-w_j)^{-1}~\psi_1$. Hence the eigenvalue equation is indeed obtained by setting the right hand side of (\ref{det}) to zero. 

With this simple form for $det(z-H)$ we obtain the potential defined in (\ref{pot}) as

\beq\label{jack}
\Phi(z,z^\ast) ~=~ {1 \over 2N} ~\lag ~log \mid \left( \Pi_{i=1}^N ~(z-w_i)\right) ~-1 \mid^2 \rag
\eeq
Let us understand the behavior of this function. Simply stated, the essence of our proof is to note that the quantity $\prod_{i=1}^N ~(z-w_i)$ in (\ref{jack}) goes suddenly from being neglibible compared to the $1$ in (\ref{jack}) to being completely dominant over the $1$. The suddenness of the transition is due to the large $N$ limit. The quantity in question is the product of $N$ factors, with $N$ tending to infinity. Thus, if each of the factors is larger than $1$ on average this quantity is exponentially large, and if each of the factors is smaller than $1$ on average, exponentially small. 

We can flesh out this argument somewhat. Consider the absolute magnitude of $\Pi_{i=1}^N ~ (z-w_i)$ and ask if it is larger or smaller than $1$ in (\ref{jack}). For convenience, define

\beqra\label{ref34}
q(z) &\equiv &{1 \over N} ~log~ \mid \prod_{i=1}^N ~(z-w_i) \mid \nonumber \\
&\equiv &{1 \over N} ~\sum^N_{i=1} ~log~ \mid z-w_i \mid
\eeqra

Rather than dress up our proof with mathematical abstractions, we will describe what actually happens in specific numerical terms. To focus our thoughts, let us imagine we are doing numerical computation with $N=10^3$ with $\pw$ the box distribution with $V=1$, that is, the $w_i$'s are evenly distributed between $-1$ and $+1$. Consider $q(1.5)$, say, the average of $10^3$ numbers ranging between $log~0.5$ and $log~2.5$ and equal to $log~1.5$ on the average. Obviously, $q(1.5)$ is most likely positive. In contrast, consider $q(0.5)$, the average of $10^3$ numbers ranging between $log~ \mid -0.5 \mid$ and $log~1.5$. Evidently, the function $q(x)$ is most likely negative for $x$ less than some critical value $x_\ast$ and then most likely goes positive for $x$ greater than $x_\ast$. (For a given realization $\{w_i\}$, the function $Q(z) ~\equiv~e^{Nq(z)}$ is the absolute value of a perfectly smooth polynomial with $N$ zeroes at the $w_i$'s, and does not fluctuate significantly from realization to realization since for $N$ large, realizations in which most of the $w_i$'s are almost equal to $+1$, say, are exponentially unlikely.) In the $N \rta \infty$ limit, the function $Q(x)$ jumps from zero for $x~<~x_\ast$ to infinity for $x~>~x_\ast$. To see that this jump is abrupt in the large $N$ limit, consider that in going from $N~=~10^3$ to $N~=~10^6$, we basically take the $Q(x)$ for $N~=~10^3$ and raise it to the power $10^3$. Similarly, for a fixed $\theta$, as we increase $r$ there is a critical $r_\ast(\theta)$ at which $Q(re^{i\theta})$ jumps from zero to infinity.

From (\ref{jack}) we see that for a given $N$ and for a given realization, $\Phi(z,z^\ast)$ goes from essentially zero (when $Q(z)~=~\mid \prod_{i=1}^N ~(z-w_i) \mid$ is small compared to $1$) to a non-zero value (when $Q(z)$ is large and the $1$ inside the logarithm is (\ref{jack}) can be neglected.) The transition occurs when $Q(z) ~=~ 1$ and hence $q(z)=0$. As $N \rta \infty$, the transition region becomes narrower and narrower and so the transition region is one-dimensional.

Hence we see that in the complex plane there is a curve across which $\Phi(z,z^\ast)$ changes from

\beq\label{susan2}
\Phi(z,z^\ast) ~=~ 0
\eeq
to

\beqra\label{mary}
\Phi(z,z^\ast) &= &\lag {1 \over 2N}~ log \mid \Pi_{i=1}^N ~(z-w_i) \mid^2 \rag \nonumber \\
&= &\lag ~log \mid z-w \mid \rag
\eeqra

We have proved that the spectrum in the maximally non-hermitean limit is one-dimensional and that the spectral curve is given by the dividing line between (\ref{susan2}) and (\ref{mary}):

\beq\label{central}
\lag ~log~ \mid z-w \mid \rag ~=~ 0
\eeq

Our proof can be made more mathematically rigorous by the usual considerations that for $N$ large the properties of $H$ for a given realization $\{w_i\}$ are the same as the averaged properties of $H$ for $N=\infty$.  

To summarize, we want to emphasize how simple our proof is. First, we note that in the maximally non-hermitean limit the determinant of $(z-H)$ is given simply by (\ref{det}). Then we just look at (\ref{det}) and ask which of the two terms on its right hand side is larger. The transition between the two regimes gives the spectral curve.

We are now in a position to understand the wings in an extraordinarily simple fashion. In the region where (\ref{mary}) holds the density of eigenvalues is given, according to (\ref{dens}) by simply

\beq\label{wing}
\rho_{wing} ~(x,y) ~=~ \delta(y)~ \theta(x-x_\ast) P(x)
\eeq
Here $x_\ast$ is defined by the region where (\ref{mary}) holds. We see that the wings stick out from the central ``elliptical" region, since inside the curve defined by (\ref{central}) the potential $\Phi$ vanishes and hence $\rho$ vanishes.

This discussion also shows that the density of eigenvalues on the wings simply follows the probability distribution of the site energies. We can check this result against our explicit result in (\ref{N}). In the one way limit, $t ~\rta~ 0$ and hence $B ~\rta~x^2 ~+~ \gam^2$. We see that (\ref{N}) indeed reduces to (\ref{wing}). We also understand why there are no wings in the sign model in the one way limit.

While (\ref{wing}) is not immediately obvious from the Schr\"odinger equation (\ref{sch}), we can see heuristically from (\ref{det}) that for a given $w_k$ how one might be able to find an eigenvalue near $w_k$ by writing $E ~=~ w_k ~+~ \delta_k$ for $z$ in (\ref{det}). The condition for a solution is that the product $\prod_{i \ne k}(w_k - w_i)$ is approximately equal to $1 \over \delta_k$. According to (\ref{wing}), this is possible in a statistical sense for $w_k$ large enough (larger than $x_\ast$.) We see from (\ref{sch}) that the solution is clearly peaked on the $k^{th}$ site.

The reader who has read FZ and BZ may recognize that the way we derived (\ref{central}) was essentially how the spectral curve (\ref{curve}) for the sign model was obtained. In the sign model, each $w_i$ is equal to either $+r$ or $-r$ with equal probability. For $N$ large, in a given realization, most likely about half of the $w_i$'s are equal to $+r$, and the rest equal to $-r$. Thus $Q(z) ~=~ \mid \Pi_{i=1}^N ~(z-w_i) \mid ~\simeq~ \mid z^2-r^2 \mid^{{N \over 2}}$. Thus, for $\mid z^2-r^2 \mid ~>~ 1$ the $1$ in (\ref{jack}) is completely overwhelmed, while for $\mid z^2-r^2 \mid ~<~ 1$, the $1$ dominates. The spectral curve is given by $\mid z^2-r^2 \mid ~=~ 1$ as in (\ref{curve}).

We also mention that the bounds in BZ, alluded to in section I, were obtained by a related argument based on whether $\mid ~\lag\left({1 \over z-w} \right)^l \rag \mid$ is less than $1$ for some integer $l$. 

The proof of Goldsheid and Khoruzhenko is more involved but also more general, and applies to the non-maximally non-hermitean limit. In the appendix we give a much simplified version of their proof, purging what we do not need for our purposes.

Finally, the discussion in this section enables us to calculate the localization length $L~(E)$ on the wings. According to the discussion following (\ref{sch}), if we move $n$ lattice spacings away from the site where the wave function attains its peak value, the absolute value of the wave function $\mid \psi \mid$ decreases by a factor

\beqra\label{fac}
{1 \over \mid \Pi ~(E-w_j) \mid} &= &e^{- \Sigma ~log ~ \mid E-w_j \mid} \nonumber \\
&= &e^{- n ~({1 \over n}~\Sigma ~log ~ \mid E-w_j \mid )}
\eeqra
where the product and sum in (\ref{fac}) involve $n$ terms. For $n$ large, we see that the localization length $L(E)$ is none other than the inverse of the function $q(E)$ defined in (\ref{ref34}):

\beq\label{lolen}
L~(E) ~=~ {1 \over q(E)}
\eeq
Indeed, $q(E)$ is positive on the wings. As $E$ approaches $E_\ast$ (also called $x_\ast$ in some contexts) at which the wings are attached to the arcs of complex eigenvalues, $q(E)$ vanishes according to (\ref{susan2}) and hence generically

\beq\label{gen}
L(E) ~\sim~ {1 \over \mid E-E_\ast \mid}
\eeq
as we conjectured.
\pagebreak
\section{Critical Points in Maximally Non-Hermitean Models}

We will try to extract as much as we can from (\ref{central}) for a general probability distribution $P(w)$ of the site energies. (For simplicity, we will, as before, take $P(w)$ to be even.) Later, we will obtain explicit results for specific choices of $P(w)$. The equation (\ref{central}) which we write as

\beq\label{bas2}
\lag~log\left(\left(x-w\right)^2 ~+~ y^2\right) ~\rag ~=~ 0
\eeq
defines a curve $s(x,y)~=~0$ in the complex plane. With $\pw$ even, we have $s(x,y) ~=~ s(-x, y)$ and so we will henceforth focus on the first quadrant $x \geq 0~,~y \geq 0$.

Let us focus on the point $y_\ast$ at which the spectral curve $s$ intersects the imaginary axis. In general, the probability distribution $\pw$ is characterized by one or more parameters $\gam$ which measure the randomness. We will by convention take $\gam~=~0$ to describe no randomness, and associate increasing randomness with increasing $\gam$. Define

\beq\label{f}
f(y,\gam) ~=~ \lag ~log~(y^2 + w^2) ~\rag
\eeq
The desired function $y_\ast(\gam)$ is determined by 

\beq\label{y}
f(y_\ast (\gam),~\gam) ~=~ 0
\eeq
We know that with no randomness $y_\ast(0) ~=~1$, and we expect that as $\gam$ increases, $y_\ast(\gam)$ decreases until a critical point $\gam_c$ at which point $y_\ast(\gam_c)~=~0$.

These expectations are easy to prove. First, note 

\beq\label{note}
{df \over dy} ~=~ 2y \lag {1 \over y^2+w^2} \rag ~> ~0
\eeq 
(recall that we restrict ourselves to the first quadrant) and hence $f(y, \gam)$ is a monotonically increasing function of $y$. At $y=0$, the function starts with the value $f(0,\gam) ~=~ \lag ~log ~w^2 \rag$, and increases to $f ~\sim~log ~ y^2~$ for large $y$. Thus, if $\lag ~log~w^2 ~ \rag ~<~0$, the function $f(y,\gam)$ vanishes at some $y_\ast(\gam)$. But if $\lag ~log~w^2 ~ \rag ~>~0$, there is no $y_\ast(\gam)$. Hence, the critical value $\gam_c$ is determined simply by 

\beq\label{crit}
\lag~log~w^2~\rag ~=~ 0
\eeq
A critical transition occurs when the average of the logarithm of the site energy squared vanishes.

Let us give a specific example, the box distribution in (\ref{box}). Note for our purposes an overall multiplicative factor in $f(y,\gam)$ is irrelevant and thus we can just as well write

\beqra\label{bob}
f(y,\gam) &= &\int_0^V ~dw ~log~(y^2+w^2) \nonumber \\
&= &2y~ {\rm arc~ tan}~ {V \over y} ~+~ V\left(-2~+~log(y^2+V^2)\right)
\eeqra
(Here the generic parameter $\gam$ is denoted by $V$.) In particular, up to an irrelevant overall factor, we may write (\ref{crit}) as

\beq\label{bill}
-2~+~log~V_c^2 ~=~ 0
\eeq
and thus we find

\beq\label{vc}
V_c~=~e
\eeq

As we increase the randomness (as measure by $\gam$), we expect the eigenvalues to migrate towards the real axis, and hence $y_\ast(\gam)$ to decrease. This expectation can be easily proved for a general $\pw$. To avoid notational clutter, let $\pw$ contain only one parameter $\gam$ relative to which the random site energy $w$ is measured so that we can write $\pw ~=~ c(\gam) ~p({w \over \gam})$. (Thus, for example, for the box distribution (\ref{box}), the one parameter $\gam$ is called $V$, and $c(\gam) ~=~ {1 \over 2V}$, and $p({w \over \gam}) ~=~ \theta\left(1~-~({w \over \gam})^2\right)$.) Differentiating the defining equation for $y_\ast(\gam)$

\beq\label{wom}
\int_0^{\infty}~dw~p({w \over \gam}) ~ log ~ (y_\ast^2 ~+~ w^2) ~=~ 0
\eeq
with respect to $\gam$ and integrating by parts, we obtain

\beq\label{mat}
\int_0^{\infty}~dw ~ \left(y_\ast~{dy_\ast \over d \gam} ~+~ {w^2 \over \gam}\right) ~p\left({w \over \gam}\right)~\left({1 \over w^2+y^2_\ast}\right) ~=~ 0
\eeq
thus proving that ${dy_\ast \over d\gam}~<~0$, as desired.

Indeed, it is easy to obtain a general result for small randomness. Expanding (\ref{y})

\beq\label{yy}
\lag ~log~ \left(y^2_\ast ~+~w^2 \right) \rag ~=~ 0
\eeq 
 for $y_\ast ~=~ 1-\delta y$, we obtain $\lag -2\delta y ~+~ w^2 \rag ~=~ 0$ and hence immediately

\beq\label{yyy}
y_\ast (\gam) ~=~ 1 ~-~ {1 \over 2} \lag w^2 \rag + \cdots 
\eeq 
for small randomness. In particular, for the box distribution, $y_\ast (V) ~=~ 1 - {1 \over 6} ~V^2 + \cdots$. Also, for the box distribution, in the other extreme, when randomness reaches the critical value $V_c$ given in (\ref{vc}), $y_\ast(V) ~\simeq~ {2 \over \pi}(V_c-V)$ vanishes linearly.

It is now just a matter of doing the appropriate integral to obtain $\gamma_c$ for various probability distributions of the reader's choice. Notice that in many cases we may be able to determine $\gam_c$ analytically without being able to determine $y(\gam)$ analytically. Let us give some examples. The simplest is the sign distribution (\ref{pw1}) for which $f(y,r) ~=~ log ~(y^2 + r^2)$ and thus

\beq\label{sign}
y_\ast ~=~ \sqrt{1-r^2}
\eeq
giving $r_c=1$ in agreement with (\ref{curve}). It is also interesting to look at the Cauchy distribution (\ref{Cauchy}) for which we find

\beq\label{cau}
\int_0^\infty ~dw ~ {log~w^2 \over w^2 ~+~ \gam^2} ~=~ {\pi \over \gam} ~log ~\gam
\eeq
and thus

\beq\label{cau2}
\gam_c ~=~ 1
\eeq
in agreement with (\ref{critical}).

A class of probability distributions for which $\lag log ~w^2 \rag$ is easily evaluated is the generalization of the box distribution defined by $\pw ~ \propto ~ \mid w \mid^k ~ \theta \left(V^2-w^2\right)$. We obtain

\beq\label{k}
V_c ~=~ e^{{1 \over k+1}}
\eeq
and thus recover (\ref{vc}) for $k=0$. Numerically, we have $V_c ~=~ (2.718,1.649, 1.395, \cdots)$ for $k~=~(0,1,2, \cdots)$. The general trend agrees with our intuition: for a given $V$, as $k$ increases, the random site energies become larger, and hence $V_c$ should decrease. As $k ~\rightarrow~ \infty$ we have $V_c=1$, in agreement with $r_c=1$ for the sign model as expected. Similarly, we can study the ``parabolic" distribution $\pw ~ \propto ~ \left(V^2-w^2\right)~\theta(V^2-w^2)$ and distributions similar to it. We find

\beq\label{four}
V_c ~=~ e^{4 \over 3} ~=~ 3.794
\eeq

For the Gaussian distribution $\pw ~ \propto ~ e^{-{w^2 \over a^2}}$ we obtain

\beq\label{euler}
a_c ~=~ 2e^{{E \over 2}} ~=~ 2.669
\eeq
where $E$ denotes Euler's constant $0.577$. For the exponential distribution $\pw ~\propto ~ e^{-{\mid w \mid \over a}}$ we obtain

\beq\label{exp}
a_c ~=~ e^E~=~1.781
\eeq
For the same $a$, the exponential distribution is ``more random" than the Gaussian distribution, and hence we expect $a_c$ to be smaller for the exponential than for the Gaussian distribution.

We have obtained the value of $\gam_c$, the critical measure of randomness, for a variety of probability distributions $\pw$. What happens when the randomness reaches its critical strength? In the Cauchy model, we know that at $\gam_c$ all eigenvalues become real and the extended states disappear. In general, with increasing randomness, eigenvalues tend to drift towards the real axis, and thus we might guess that at $\gam_c$ all eigenvalues become real, as in the Cauchy model. This however is not true in general. In fact, we have already encountered a counter-example, the sign model. According to (\ref{curve}), for $r>r_c ~=~ 1$, the spectrum splits into two disjoint lobes. It is indeed true, however, that $y_\ast(r)$ goes to zero as $r$ increases towards $r_c$, as indicated by (\ref{sign}).

This instructive example tells us that to determine whether the eigenvalues have all collapsed onto the real axis, it is not enough to study the intersection of the spectral curve with the imaginary axis. We have to study the intersection of the spectral curve with the real axis.

According to (\ref{bas2}), the point $x_\ast(\gam)$ at which the spectral curve intersects the real axis is determined by

\beq\label{xstar}
\lag ~log ~(x_\ast-w)^2 \rag ~=~0
\eeq

%%%%%%%%%%%%%%%%%ADDITION%%%%%%%%%%%%%%%%%%%%

With $\pw$ even, we can symmetrize in $w$ and define

\beq\label{gx}
g(x,\gam) ~=~ \lag ~log(x^2 - w^2)^2 \rag
\eeq
From (\ref{bas2}) directly, or from (\ref{f}), we see that $g(x,\gam)$ is given by the analytic continuation of $2f(y,\gam)$. The sign flip between $f(y,\gam)$ and $g(x,\gam)$ makes a difference of course: it is no longer true that $g(x,\gam)$ is a monotonically increasing function of $x$. 

However, we will now prove that if $\pw$ does not increase for $w$ positive and increasing, then $g(x,\gam)$ is indeed monotonically increasing as $x$ increases. The proof is straightforward. Using $\pw ~\geq~ P(x)$ for $w<x$ and $-\pw ~\geq~ -P(x)$ for $w>x$, we find that 

\beq\label{mono}
{1 \over 4x}~{\partial g(x,\gam) \over \partial x} ~\geq~ P(x) \int_0^\infty ~ dw~ {1 \over x^2-w^2} ~=~0
\eeq
(where the last integral is understood in the principal value sense, of course.) For large $x$, $~g(x,\gam) ~\sim~ log ~x^4$. Thus, if $g(0,\gam) ~=~ \lag log ~ w^4 \rag$ is negative there exists a solution $x_\ast ~(\gam)$ to the equation

\beq\label{starx}
\lag ~log~(x_\ast^2 ~-~w^2)^2 \rag ~=~ 0
\eeq
When $g(0,\gam)$ goes positive, there is no solution, and thus the critical value of randomness is given by

\beq\label{critgam}
\lag ~log~w^2 \rag ~=~ 0
\eeq 
in agreement with (\ref{crit}), as it should.

The behavior of $x_\ast(\gam)$ for small randomness is also easy to extract. Writing $x_\ast(\gam) ~=~ 1 + \delta x$ and expanding (\ref{starx}), we find immediately that

\beq\label{small}
x_\ast(\gam) ~=~ 1 ~+~ {1 \over 2} ~\lag w^2 \rag + \ldots
\eeq
(This result and (\ref{yy}) are of course in agreement with the appendix in FZ.) That $x_\ast(\gam)$ vanishes as $\gam ~\rta~ \gam_c$ imply that $x_\ast(\gam)$ has a maximum at some critical value $\gam_\ast$.
\begin{figure}[htbp]
\epsfxsize=4in
\begin{center}
\leavevmode
\epsfbox{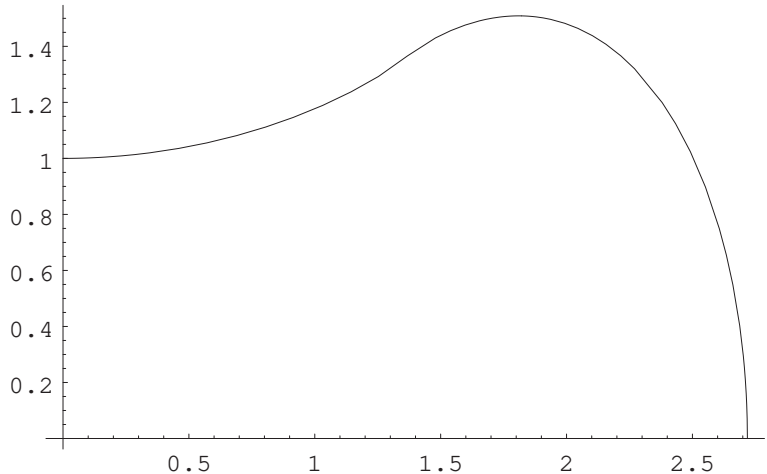}
\end{center}
\caption{The energy at which the localization-delocalization transition occurs, $x_\ast(V)$, as a funtion of $V$ for the box distribution. It has a maximum at $V_\ast ~\sim~ 1.81$ and vanishes at $V_c = e$.}
\label{fig3}
\end{figure} 

We show the behavior of $x_\ast(\gam)$ as $\gam$ varies for the box distribution in fig(3). We can evaluate $g(x,V)$ exactly to find (up to irrelevant overall factors)

\beq\label{boxg}
g(x,V) ~=~ (V-x)~log~(V-x)^2~+~(V+x)~log~(V+x)^2~-~4V
\eeq 
(where as before we remind the reader that the generic parameter $\gam$ is called $V$ for the box distribution.) To determine $\gam_\ast$ (that is, $V_\ast$ in this context), we simply differentiate $g(x,V)$ to obtain

\beq\label{stars}
V_\ast^2 ~-~ x_\ast^2(V_\ast) ~=~ 1
\eeq
Inserting this into $g(x_\ast(V_\ast), V_\ast) ~=~ 0$ we find readily that $V_\ast ~=~ cosh ~ \theta_\ast ~\simeq~ 1.81$ and $x_\ast(V_\ast) ~\simeq~ 1.51$ where $\theta_\ast$ is the solution of the equation $\theta_\ast ~tanh~\theta_\ast ~=~ 1$.

For $V \sim V_c ~=~ e$ we find $x_\ast(V) ~\simeq~ \sqrt{2V_c ~ (V_c - V)}$. We can evidently define a variety of critical exponents here: for example, 

\beq\label{beta}
x_\ast ~(V) ~\simeq~ (V_c-V)^\beta
\eeq

The alert reader may have noticed that for the Cauchy distribution $x_\ast(\gam) ~=~ \sqrt{1-\gam^2}$ and instead of increasing with $\gam$ as shown in (\ref{small}) $x_\ast(\gam)$ decreases with $\gam$. This is in fact consistent with (\ref{small}) since the second moment $\lag w^2 \rag$ does not exist for the Cauchy distribution. The critical exponent $\beta$ again takes the generic value $\beta ~=~ {1 \over 2}$.

For a $\pw$ which increases as $w$ positive increases, the behavior we just sketched for $x_\ast(\gam)$ fails. Our counter-example is the sign model, in which

\beq\label{signg}
g(x,r) ~=~ log~(x^2-r^2)^2
\eeq
a function which starts at $log ~r^4$, goes to $-\infty$ to $x~=~r$, and then rises from $-\infty$. Thus, at $r_c ~=~ 1$ we go from a single solution for $x_\ast(\gam)$ to two. (As said before, we always work in the first quadrant.) This is consistent with (\ref{curve}).

In summary, for $\pw$ which does not increase as $w$ positive increases and which has a second moment, $x_\ast(\gam)$ increases as the randomness $\gam$ increases, reach a maximum at $\gam_\ast$, and then decreases towards zero as $\gam~\rta~\gam_c$. Meanwhile, $y_\ast (\gam)$ decreases monotonically towards zero as $\gam$ increases towards $\gam_c$. Note that $x_\ast(\gam)$ has the important physical meaning as the critical energy $E_\ast(\gam)$ at which the localization-delocalization transition occurs for a given measure of randomness $\gam$.

Let us use the terminology the ``arc part" to refer to the spectrum of $H$ consisting of complex eigenvalues. Thus, increasing randomness squashes the arc part towards the real axis, elongating it in the $x$ direction, until the critical randomness $\gam_\ast$, after which it shrinks steadily, giving up spectral weight to the wings, until it disappears altogether at $\gam_c$, at which point all states become localized.

Using the sort of techniques developed here, the reader can readily deduce further properties of the spectral curve. For instance, with $\pw$ even and writing $z ~=~ re^{i\theta}$, one has

\beq\label{tom}
\lag ~ log ~\mid (z-w)(z+w) \mid ~ \rag ~=~ \lag log \left(r^4-2~r^2 ~w^2 ~cos ~2\theta + w^4 \right) ~\rag ~=~ 0
\eeq
For $\theta ~>~ {\pi \over 4}$ in the first quadrant, $cos ~2\theta$ is negative, and one can deduce that ${dr \over d\theta} ~<~ 0$ upon differentiating (\ref{tom}).

\pagebreak
\section{Appearance and Disappearance of Wings}

Starting with the spectrum with no disorder present, we expect that as we increase the randomness wings appear at some critical value of the randomness $\gam_{c1}$, as defined earlier. For the Cauchy distribution we saw explicitly that $\gam_{c1} ~=~ 0^+$; in other words, wings appear as soon as the random site energies are switched on. As we will see, this is due to the fact that the Cauchy distribution has infinitely long tails.

In this section we consider $\pw$ with support up to $w_{max} ~=~ \gam$. (For example, for the box distribution the maximum value of $w$ is $V$, which is what we denote the generic parameter $\gam$ by for this distribution.) According to (\ref{wing}), the right wing extends out to $w_{max} ~=~ \gam$. On the other hand, the right wing starts at $x_\ast(\gam)$. Clearly, the wings disappear when $x_\ast(\gam)$ becomes larger than $\gam$. Thus, we conclude that for this class of distributions $\gam_{c1}$ is determined by 

\beq\label{gamc}
x_\ast ~(\gam_{c1}) ~=~ \gam_{c1}
\eeq
Again assuming as before that $\pw$ contains only one parameter $\gam$, we see that $\gam_{c1}$ is the solution of

\beq\label{gaga}
\int_0^\gam ~ dw ~ p \left({w \over \gam}\right) ~ log (\gam^2 - w^2)^2 ~=~ 0
\eeq
 Changing variable in (\ref{gaga}) by $w ~=~ \gam u$, we obtain the result

\beq\label{gaba}
log ~ \gam_{c1}^2 ~=~ - ~ {\int_0^1 ~du~p(u)~log(1-u^2) \over \int_0^1 ~du~p(u)} ~=~ -\lag ~log(1-u^2) \rag
\eeq
which is to be compared to 

\beq\label{gaca}
log ~ \gam_{c}^2 ~=~ - ~ {\int_0^1 ~du~p(u)~log~ u^2 \over \int_0^1 ~du~p(u)} ~=~ -\lag ~log ~u^2 \rag
\eeq

In particular, for the box distribution we have

\beq\label{bxx}
V_{c1} ~=~ {1 \over 2} e
\eeq
compared to 

\beq\label{vcee}
V_c ~=~ e
\eeq
For the ``parabolic" distribution mentioned earlier we have

\beq\label{par}
V_{c1} ~=~ {1 \over 2} e^{5 \over 6}
\eeq
compared to

\beq\label{parc}
V_c ~=~ e^{4 \over 3}
\eeq

Notice that the argument shows that if $w_{max} ~=~ \infty$, as is the case for the Cauchy distribution, then the right wing extends out to infinity, and the wings are always present as long as $\gam > 0$. Thus, $\gam_{c1} ~=~ 0^+$ as we have already learned for the Cauchy distribution. Physically, for $\pw$ with $w_{max} ~=~ \infty$ the site energies have a non-zero probability of attaining arbitrarily large values and a small amount of randomness can cause the wings to appear.

It is easy to show that if $\pw$ does not increase for $w$ positive and increasing, then $\gam_{c1}^2 ~<~ \gam_c^2$. Start by noting that $(1-u)^2 ~\leq~ (1-u^2)$ for $0 ~\leq~u~\leq~1$, and that for the class of $\pw$'s under consideration $\lag - ~log ~u^2 \rag ~\geq~ \lag - ~log ~(1-u)^2 \rag$. Combining these two inequalities we find the desired result.

\pagebreak
\section{Spectral Curve for the Box Distribution}

In the preceding section, we studied the points $x_\ast(\gam)$ and $y_\ast(\gam)$ at which the spectral curve $s(x,y)~=~0$ intersects the real and imaginary axes for general distributions $\pw$ of the site energies. In this section, we determine the entire spectral curve $s(x,y)~=~0$ for the box distribution (\ref{box}).
 
First, we evaluate (\ref{mary}) to obtain

\beq\label{bxx}
\Phi(z ~=~ Vu,V) ~=~ {1 \over 2}~\left(log ~ \mid {(u+1)^{u+1} \over (u-1)^{u-1}}~V^2 \mid \right)~-~ 1
\eeq
where we scale $z ~=~Vu$ for convenience. Thus, the spectral curve is determined by the interesting equation

\beq\label{spe}
\mid {(u+1)^{u+1} \over (u-1)^{u-1}} \mid ~=~ {e^2 \over V^2}
\eeq
Let us write $u~=~s~+~it$ and define the left hand side of (\ref{spe}) as $b(s,t)$. We find

\beq\label{defb}
b(s,t) ~=~ \left( (s+1)^2 ~+t^2 \right)^{{1+s \over 2}}~\left( (s-1)^2 ~+t^2 \right)^{{1-s \over 2}} ~e^{-t ~arc~tan~{2t \over 1-s^2-t^2}}
\eeq
The spectral curve (with $z ~=~ x+iy$ ) is then defined by

\beq\label{bsc}
b({x \over V},~{y \over V}) ~=~ {e^2 \over V^2}
\eeq

In particular, $x_\ast(\gam)$, the critical energy at which localization-delocalization transition occurs, is determined by the remarkable function

\beq\label{rem}
b(s,0) ~=~ (s^2-1) ~ ({s+1 \over s-1})^s
\eeq
In solving (\ref{bsc}) it is essential to pick the correct branch of the $arc ~tan$ function, which is defined up to multiples of $\pi$. Let us take the conventional definition of $arc~ tan$ as a function ranging between $-{\pi \over 2}$ and ${\pi \over 2}$. Then we see that for $t^2 ~<~ 1-s^2$, $b(s,t)$ has to be multiplied by a factor of $e^{\pi t}$. This ensures that $b(s,t)$ is continuous as $t$ passes through $\sqrt{1-s^2}$. We show in figure (4) the solution of (\ref{bsc}) for $V=1$. The spectral curve is ``elliptical" in shape. We find numerically that as $V$ approaches $V_c = e$ the fluctuation between runs even for $N=200$ to be quite significant. These fluctuations are described by the density-denstity correlation function discussed in BZ. 

We showed in figure (3) the behavior of the energy of the localization-delocalization transition $x_\ast(V)$, defined by 

\begin{figure}
\epsfxsize=4in
\begin{center}
\leavevmode
\epsfbox{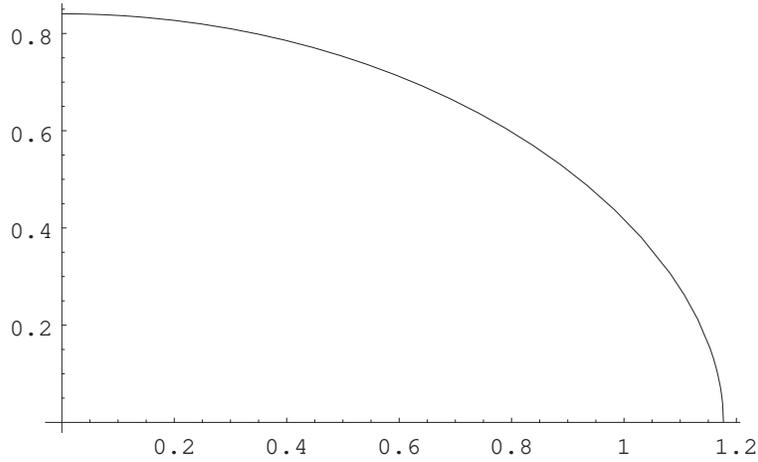}
\end{center}
\caption{Spectral Curve for the box distribution with $V=1$, plotted in the first quadrant}
\label{fig4}
\end{figure}

\beq\label{bsc2}
b({x_\ast \over V}, 0) ~=~ {e^2 \over V^2}
\eeq
The function $x_\ast(V)$ behaves as discussed in the last section. In particular, at $V=1$ the solution of 

\beq\label{rem1}
(x^2 ~-~ 1)~({x+1 \over x-1})^x ~=~ e^2
\eeq
is given by $x=1.17697$. Note also that $b(1,0) ~=~ 4$ and so for $V ~=~ {e \over 2}$, we have $x_\ast(V) ~=~ V$. Recall from the preceding section that $V_{c1}$ for this model is precisely ${e \over 2}$. We see that indeed the wings disappear at this critical value of the randomness. 

Given the complexity of (\ref{bsc}) it would be a challenge to determine the spectral curve in the non-maximally non-hermitean case.

\pagebreak
\section{Dilution}

As mentioned in the Introduction, the effects of dilution were considered in BZ. We point out here that the effects of dilution can be quite drastic. The sign model is most amenable to analytic study with (\ref{dil})

\beq\label{dilsgn}
P_d(w) ~=~ d ~\delta(w) ~+~ {1 \over 2}(1-d) ~(\delta(w-r) ~+~ \delta(w+r))
\eeq

The spectral curve is determined by

\beq\label{delspe}
(x^2 + y^2)^d ~\left[\left(x^2-y^2-r^2\right)^2~+~4x^2y^2\right]^{{1-d \over 2}} ~=~ 1
\eeq
Consider for simplicity the equation for $y_\ast(r)$. For no dilution $d=0$, this becomes

\beq\label{nodil}
y_\ast^2 ~+~ r^2 ~=~ 1
\eeq
as in (\ref{beta}). For any amount of dilution, $d>0$, we have

\beq\label{dill}
y_\ast^{2d}(y_\ast^2 ~+~r^2)^{1-d} ~=~ 1
\eeq
Note the left hand side of both (\ref{nodil}) and (\ref{dill}) are monotonic in $y_\ast$, but the left hand side of (\ref{dill}) starts at $0$ instead of $r^2$ and so a solution of (\ref{dill}) for $y_\ast$ always exists for any $d>0$. This is in sharp contrast with (\ref{nodil}), for which $y_\ast$ exists only for $r<1$. As shown in fig(5), for $r~>~1$ and any $d ~>~ 0$ a new ``lobe" centered at the origin appears in the spectrum, with this lobe intersecting the imaginary axis at $y_\ast(r,d)$. We can easily solve (\ref{dill}) for $d={1\over2}$, obtaining

\beq\label{jill}
y_\ast(r,d ={1 \over 2}) ~=~ \sqrt{{1 \over 2}(\sqrt{r^4 + 4}~-~ r^2)}
\eeq
More generally, for any $d$, we see that for small $r$,

\beq\label{j2}
y_\ast(r,d) ~\rta~ {1 \over 2}~(1-d)~ r^2 +\cdots
\eeq
and for large $r$

\beq\label{j3}
y_\ast(r,d) ~\rta~{1 \over r^{{(1-d) \over d}}}~+\cdots
\eeq
For intermediate $r$, $y_\ast(r,d)$ interpolates smoothly between (\ref{j2}) and (\ref{j3}).
\begin{figure}[htbp]
\epsfxsize=4in
\begin{center}
\leavevmode
\epsfbox{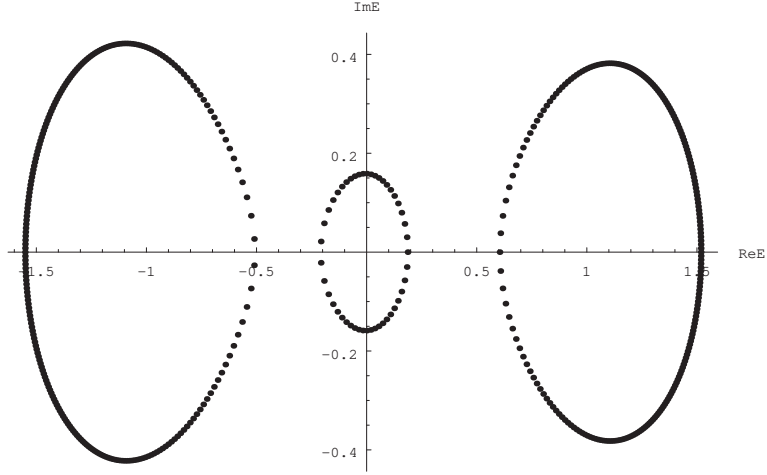}
\end{center}
\caption{The diluted sign model with $d=0.1$ and $r=1.2$}
\label{fig5}
\end{figure}

We see that there is a critical transition with $d_c =0$, as emphasized in particular by (\ref{j3}).

More generally, recalling the definition of $f(y, \gam) ~=~ \lag ~log~(y^2 + w^2)\rag$ from (\ref{f}) associated with $\pw$, we have associated with $P_d(w)$ the function

\beq\label{fyd}
f(y,\gam,d) ~=~ d ~log~ y^2 ~+~(1-d) ~f(y,\gam)
\eeq
with the intersection of the spectral curve with the imaginary axis $y_\ast(\gam,d)$ given by $f(y_\ast,\gam,d) =0$ as in (\ref{y}). Evidently, as the dilution goes to its maximum value $d \rta 1$, $y_\ast \rta 1$.

For the Cauchy distribution, we have

\beq\label{cauf}
f(y,\gam) ~=~ log ~(y+\gam)^2
\eeq
With no dilution, $y_\ast(\gam) ~=~ 1-\gam$ as in FZ. For small $d$,

\beq\label{smalld}
y_\ast(\gam,d) ~=~ 1 - \gam ~-~ (log(1-\gam))~ d ~+~ \cdots
\eeq
Clearly as $\gam$ nears its critical value $1$, a small dilution has a big effect. For $d \stackrel{{\textstyle<}}{\sim} ~1$,

\beq\label{dnear1}
y_\ast(\gam,d) ~=~ 1 ~-~ (~log~(1+\gam)) ~(1-d)~+~\cdots
\eeq
Further study of the effects of dilution would be interesting.

{\bf Acknowledgements}~~~
This work was partly supported by the National Science Foundation under Grant No. PHY89-04035.

\pagebreak
\section{\sloppy Appendix: The Proof of Goldsheid and Khoruzhenk\-o Simplified}

Here we give a simplified version of the proof of Goldshied and Khoruzhenko. Let us go back to (\ref{H1}). The Hamiltonian $\tilde{H}$ would be hermitean had the entries $\tilde{H}_{N1}$ and $\tilde{H}_{1N}$ been absent. It is natural to define the hermitean Hamiltonian $H_h$ by

\beq\label{Hh}
\tilde{H} ~=~ H_h +V
\eeq
where all entries of $V$ are zero except 
\beqra\label{v}
V_{N1} &= &{t \over 2} e^{Nh} \nonumber \\
V_{1N} &= &{t \over 2} e^{-Nh}
\eeqra
(At first sight, the transformation of $H$ into $\tilde{H}$ appears to be a potentially dangerous manuever \cite{banks} as $\tilde{H}$ is an $N$ by $N$ matix with an entry which explodes exponentially as $N \rightarrow \infty$; in fact, we are going to exploit precisely this feature.) 

Since $H$ and $\tilde{H}$ have the same spectrum, we can obtain the desired density of eigenvalues of $H$ by 

\beq\label{rho}
\rho(x,y) ~=~ {2 \over \pi}~ {\partial^2 \over \partial z~ \partial z^\ast}~\tilde{\Phi}(z,z^\ast)
\eeq
with

\beq\label{F}
\tilde{\Phi}(z,z^\ast) ~\equiv~ {1 \over 2N} \lag log~ \mid det ~ (z-\tilde{H}) \mid^2 \rag
\eeq
In analogy to (\ref{F}) we also define

\beq\label{big}
\Phi_h(z,z^\ast) ~\equiv~ {1 \over 2N} \lag log~ \mid det ~ (z-H_h) \mid^2 \rag
\eeq
corresponding to the hermitean Hamiltonian $H_h$. Using

\beq\label{diff}
(z-\htil) ~=~ \left(I-VG\right) ~ (z-H_h)
\eeq
with

\beq\label{green2}
G~\equiv~{1 \over z-H_h}
\eeq
(not to be confused with the $G$ we defined in (\ref{green}); we use the same symbol to avoid notational clutter) we have immediately

\beq\label{phi}
\tilde{\Phi}(z,z^\ast) ~=~ \Phi_h(z,z^\ast)~+~{1 \over 2N} \lag log~ \mid det ~ (I-VG) \mid^2 \rag
\eeq
Since $V$ has only two non-zero entries, the determinant in (\ref{phi}) can be evaluated immediately to be

\beq\label{large}
{\rm det} ~(I-VG) ~=~ (1 ~-~{t \over 2}~e^{-Nh} G_{N1}) ~(1~-~{t \over 2}e^{Nh}G_{1N})~-~{t^2 \over 4}~G_{11}~G_{NN}
\eeq
The large $N$ behavior of $det(I-VG)$ thus depends on whether $G_{1N}$ vanishes faster than $e^{-Nh}$ or not. 

If $G_{1N}$ vanishes faster than $e^{-Nh}$, $det(I~-~VG)$ is bounded. (Since ${1 \over N}{\rm tr} G$ is the Green's function of a hermitean Hamiltonian it has a finite large $N$ limit, and $G_{11}$ and $G_{NN}$ are expected to be the same order of magnitude as ${1 \over N}{\rm tr} G$.) Then from (\ref{phi}) we have

\beq\label{same}
\tilde{\Phi}(z,z^\ast) ~=~ \Phi_h(z,z^\ast)
\eeq

In contrast, if $G_{1N}$ does not vanish faster than $e^{-Nh}$, then 

\beq\label{lim}
det~(I ~-~ VG) ~\rta~ - {t \over 2} ~e^{Nh}~ G_{1N}
\eeq
as $N\rightarrow \infty$, and we have to keep the second term in (\ref{phi}). From the definition (\ref{green2}) and the form of $H_h$ we obtain immediately

\beq\label{det2}
\mid G_{1N} \mid ~=~ {\left({t \over 2}\right)^{N-1} \over \mid ~det~ \left(z-H_h\right)\mid}
\eeq
and hence from (\ref{big})

\beqra\label{log}
\lag log\mid G_{1N}\mid \rag &= &log~\left({t \over 2}\right)^{N-1} ~-~\lag log \mid det \left(z-H_h\right)\mid \rag \nonumber \\
&= &log \left({t \over 2} \right)^{N-1} ~-~ N \Phi_h (z,z^\ast) 
\eeqra
Inserting (\ref{log}) in (\ref{lim}) we obtain

\beqra\label{woa}
{1 \over N}\lag log \mid det(I-VG)\mid \rag &= &{1 \over N} ~ \left[ log\left(\left({t\over2}\right)^N e^{Nh}\right) ~-~ N\Phi_h(z,z^\ast) \right] \nonumber \\
&= &log \left({t \over 2} e^h \right) ~-~ \Phi_h (z, z^\ast)
\eeqra
At this point we might as well set $t=2$. Thus, combining (\ref{phi}) and (\ref{woa}) we see that if $G_{1N}$ does not vanish too fast we have

\beq\label{const}
\tilde{\Phi}(z,z^\ast) ~=~ h
\eeq
According to (\ref{rho}), $\htil$ and hence $H$ do not have any eigenvalues in this region.

We thus obtain the result of GK that the dividing line between (\ref{same}) and (\ref{const}) is the curve defined by

\beq\label{cur}
\Phi_h(z,z^\ast) ~=~ h
\eeq
This forms the ``arcs" of the spectral curve of $H$. The wings are determined by the solution of (\ref{same}) which lies outside the curve defined by (\ref{cur}).

The defining equation (\ref{cur}) for the spectral curve, while elegantly compact, cannot be explicitly solved in general. We need an exact analytic expression for $\Phi_h(z,z^\ast)$ which amounts to an exact determination of the density of eigenvalues of the hermitean Hamiltonian $H_h$. To the best of our knowledge, this is available only for the Cauchy distribution(\ref{Cauchy}). Indeed, Goldsheid and Khoruzhenko \cite{gk} are then led to the curve in (\ref{M}), which was derived in BZ by a different method.

Thus, in order to obtain explicit expressions for the spectral curves for various $\pw$'s, we are compelled to got to the maximally non-hermitean or one way limit (\ref{thlim}). Going back to (\ref{woa}) we see that in this limit the spectral curve is actually defined by setting $h=0$ in (\ref{cur}).

Next we observe another drastic simplification. In this limit, the hermitean Hamiltonian $H_h$ goes over to a purely diagonal matrix with the $j^{th}$ diagonal element equal to $w_j$, and thus the function $\Phi_h(z,z^\ast)$ defined in (\ref{big}) simplifies drastically to 

\beq\label{dras}
\Phi_h(z,z^\ast) ~=~ {1 \over 2}~ \lag log ~\mid z-w \mid^2 \rag 
\eeq
Thus, in the maximally non-hermitean limit, the spectral curve of the non-hermitean Hamiltonian $H$ in (\ref{eig}) is determined by the remarkably compact equation

\beq\label{bas}
\lag ~log~\mid z-w \mid^2 \rag ~=~ 0
\eeq
in complete agreement with (\ref{central}) and our simple ``one-line" proof. It is interesting to note that in this respect the non-hermitean problem is again simpler than the hermitean problem. There is no analogous simplifying limit we can take in the hermitean localization problem.

\end{document}